\documentclass[pss]{wiley2sp} 
\usepackage{amsmath}
\usepackage[none]{hyphenat}

\begin{document}

\title{Raman and XPS analyses of pristine and annealed N-doped double-walled carbon nanotubes}

\titlerunning{Raman and XPS analyses of N-doped DWCNTs}

\author{%
  Lei Shi\textsuperscript{\Ast,\textsf{\bfseries 1}},
  Markus Sauer\textsuperscript{\textsf{\bfseries 1}},
  Oleg Domanov\textsuperscript{\textsf{\bfseries 1}},
  Philip Rohringer\textsuperscript{\textsf{\bfseries 1}},
  Paola Ayala\textsuperscript{\textsf{\bfseries 1}},
  Thomas Pichler\textsuperscript{\textsf{\bfseries 1}}
}

\authorrunning{L. Shi et al.}

\mail{e-mail
    \textsf{lei.shi@univie.ac.at}, Phone: +43-0676-4456949, Fax:+43-1427751375}

\institute{%
  \textsuperscript{1}\,University of Vienna, Faculty of Physics, Strudlhofgasse 4, 1090 Vienna, Austria\\
 }

\received{XXXX, revised XXXX, accepted XXXX} 
\published{XXXX} 

\keywords{Nitrogen dopant, double-walled carbon nanotubes, annealing, Raman scattering, X-ray photoelectron spectroscopy.}

\abstract{%
%
%
%
  N-doped single/multi-walled carbon nanotubes (CNTs) were studied for long time from synthesis to properties. However, the stability of N in the CNT lattice still needs further developments. In this work, to obtain more stable N-doped CNTs, concentric double-walled (DW) CNTs with more N were synthesized using benzylamine as C and N source. In order to test the stability of N-doped DWCNTs, high-temperature annealing in vacuum was performed. By XPS and Raman spectroscopic measurements, we found that the N-doped DWCNTs are still stable under 1500 $\,^{\circ}\mathrm{C}$: the graphitic N does not change at all, the molecular N is partly removed, and the pyridinic N ratio greatly increases by more than two times. The reason could be that the N atoms from the surrounded N-contained materials combine into the CNT lattice during the annealing. Compared with the undoped DWCNTs, no Raman frequency shift was observed for the RBM, the G-band, and the G'-band of the N-doped DWCNTs.
  }
%
%
\maketitle   
\section{Introduction}
In order to tailor the band structure, the Fermi level, and the density of states of the carbon nanotubes (CNTs), nitrogen as electron donor was used to dope the pristine CNTs \cite{Gerber09PRB,Ayala10RoMP}. The N-doping changes the electronic properties of the CNTs, including the density of mobile charge carriers, hopping mobility, and work function\cite{Wiggins-Camacho09TJoPCC}. Hence, compared with the pristine CNTs, the so--called N-doped CNTs are with better N--type conductivity \cite{Wiggins-Camacho09TJoPCC}, ideal gas sensor \cite{Villalpando-Paez04CPL}, enhanced field emission properties \cite{Chun09C}, and high performance transistor \cite{Xiao05JotACS}. Although the advantage of the N-doped CNTs is very clear, the effective and controllable synthesis of N-doped CNTs with specific type of doping is still not solved.

\indent Synthesis of N-doped CNTs was studied for long time, but most of the researches focus on N-doped multi-walled CNTs (MWCNTs), partly on single-wall CNTs (SWCNTs) \cite{Villalpando-Paez06CPL,Pint11AN,Thurakitseree12C,Li14N}, and few on double-walled CNTs (DWCNTs) \cite{Chun09C,Kim05CPL,Panchakarla07AN} because of the challenge differences for the synthesis. DWCNTs have already show their advantage on both properties of SWCNTs and MWCNTs, for example, combined the excellent mechanical, electrical, and chemical properties \cite{Shen11N}. Hence, DWCNTs were attracted more attention recently by physicists, chemists, materials scientists, and engineers. Even more, the N-doped DWCNTs would be even better for the future fundamental studies as well as applications due to their improved electrical and chemical properties compared to pristine DWCNTs. Hence, synthesis of high-purity N-doped DWCNTs is one of the key points for future researches and applications.

\indent In-situ doping of nitrogen atoms in the structure of the CNTs during the growth gives much better nitrogen stabilities than the post-treatment method, for example, under thermo–chemical treatment \cite{Golberg00C} or by hydrogen nitrate \cite{Geng07JotACS}. The nitrogen source in the in-situ synthesis normally can be both ammonia gas \cite{Wiggins-Camacho09TJoPCC,Chun09C,Kim05CPL,Panchakarla07AN} or liquid contained nitrogen (N-liquid) including ethylenediamine \cite{Li14N}, acetonitrile \cite{Pint11AN,Thurakitseree12C}, pyridine \cite{Pint11AN,Panchakarla07AN}, and benzylamine \cite{Villalpando-Paez06CPL,Ayala07TJoPCC,Ayala07pss(b),Maciel09PSS(b)}. The big advantage using N-liquid is that it not only can be used as N supplier, but also as C source. Till now, high-purity N-doped DWCNTs can be only synthesized using ammonia gas, but not N-liquid.

\indent Here, we report a detailed study on the N-doped DWCNTs with small diameter by HVCVD using pure benzylamine as C source as well as N supplier. Higher doping ratio, smaller diameter, and high-purity N-doped DWCNT buckypaper make them good candidate for applications, for example, in the fields of flexible transparent conductive film \cite{Geng07JotACS} and oxygen reduction reactions \cite{Gong09S}. In addition, we also testing the stability of the N-doped DWCNTs under high temperature. More interestingly, we found out that after annealing the doping ratio increases dramatically compared with the pristine N-doped DWCNTs. Under the high temperature, the graphitic N is very stable, the ratio for the pyridinic N increases a lot due to the conversion from the N-contained materials, and the ratio for the molecular N decreases, but it is not completely removed.\\

\section{Experimental details}
N-doped DWCNTs were synthesized by high vacuum chemical vapor deposition (HVCVD).  The HVCVD system was successfully used for the growth of pristine DWCNTs with small diameter using ethanol as the C source in our group \cite{Shi15arXiv}. Here, in order to synthesize N-doped thin DWCNTs, pure benzylamine (Sigma-Aldrich, 99.9 wt.$\%$) was used as C source as well as N supplier. Ammonium iron citrate (Sigma-Aldrich, 3 wt.$\%$) was mixed with MgO (Sigma-Aldrich, 97 wt.$\%$) in ethanol, sonicated in bath for 1 hour, and dried in beaker at 70~$\,^{\circ}\mathrm{C}$. The growth was carried out at different temperatures between 825 and 900~$\,^{\circ}\mathrm{C}$ with a pressure at $\sim$1 mbar. Undoped DWCNTs were also synthesized using ethanol for comparison. Note that hydrogen, which usually was used to reduce the catalysts, is not necessary in our synthesis.

\indent After synthesis, the as-grown N-doped DWCNTs was purified by a three-step process. Firstly, the obtained powder was immersed in HCl ($\sim$37 wt.$\%$) for 2 hours to dissolve and remove most of the MgO and part of iron particles. Secondly, the sample was heated in dry air for half hour at 400~$\,^{\circ}\mathrm{C}$ to remove the amorphous carbon and the graphitic carbon surrounded the iron particles, and then immersed in HCl again for 24 hours to remove the rest of the MgO and exposed iron particles. Thirdly, the dried powder was heated in air at 500 ~$\,^{\circ}\mathrm{C}$ for 2 hours to remove the thin or destroyed SWCNTs. Finally, a thin buckypaper was formed after sonicating the obtained DWCNT powder in ethanol and filtered by membrane (MF-Millipore, 0.22 $\mu$m). Typically, 0.5 g catalyst was used for synthesis, and 30 mg N-doped DWCNTs can be obtained after purification.

\indent The purified N-doped DWCNTs were annealed in vacuum ($\sim$ 10$^{-7}$ mbar) at 1500 ~$\,^{\circ}\mathrm{C}$ to test the nitrogen stability and to improve the graphitization of the N-doped DWCNTs.

\indent The samples were measured by a Raman spectrometer (Horiba Jobin Yvon, LabRAM HR800) in ambient conditions with multi-laser lines for excitation. The slit width was set at 200 $\mu$m. Standard-resolution mode was used for the measurement by the help of a x50 objective, so the spectral resolution was $\sim$ 2 cm$^{-1}$. For ease of comparison, the spectra were all normalized by the intensity of the G-band. The nitrogen contents of the N-doped DWCNT samples were firstly annealed in vacuum (10$^{-10}$ mbar), and then probed with monochromatic Al K$\alpha$ radiation (1486.6 eV) by a hemispherical SCIENTA RS4000 photoelectron analyzer. \\

\begin{figure}[t]
\centering
\includegraphics*[width=\linewidth]{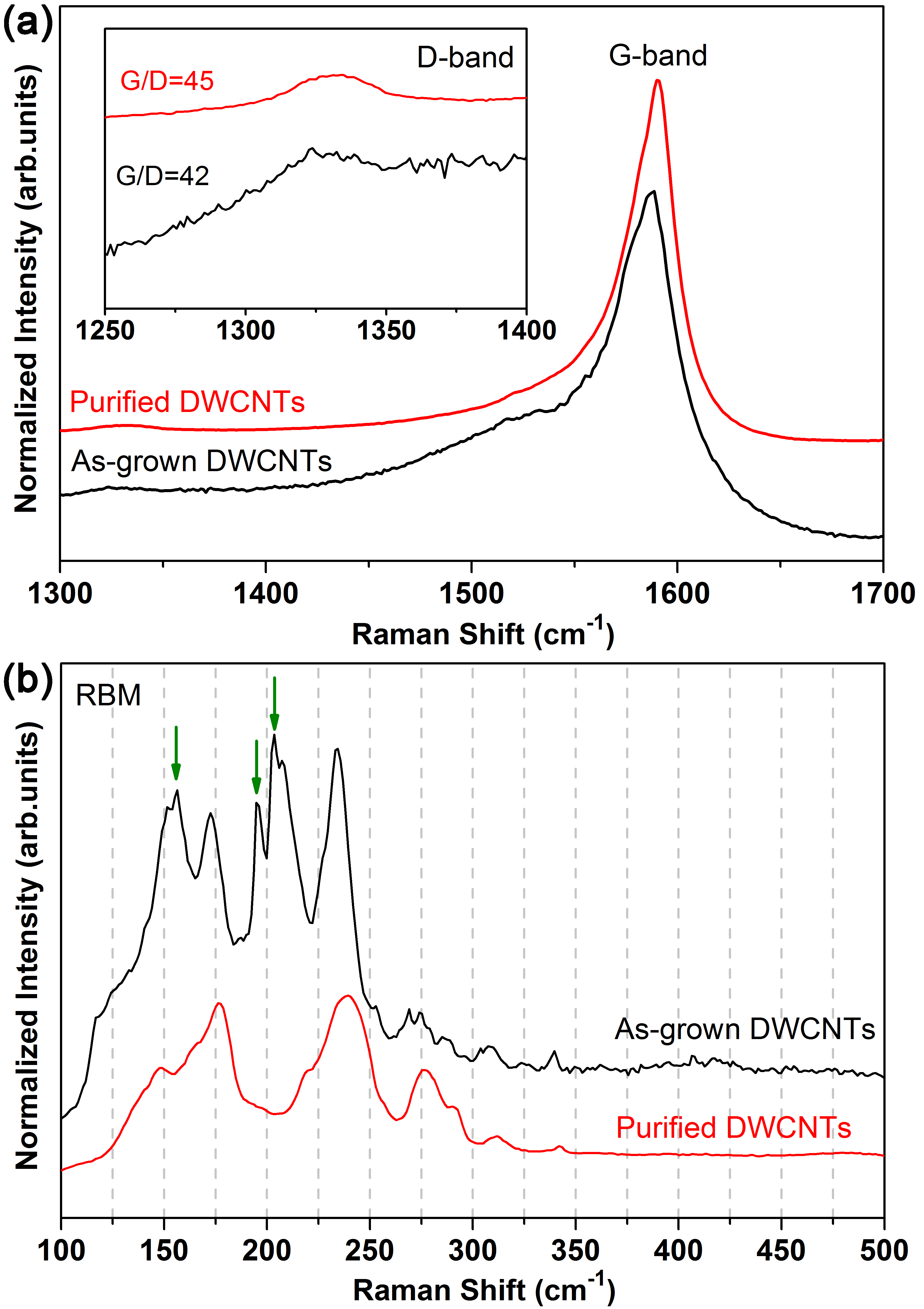}
\caption{(online color at: www.pss-b.com) Raman spectra of as-grown and purified DWCNTs excited by 568.188 nm laser. (a) G-band and D-band. Inset: A close view for the D-band. (b) RBM region.}
\label{Fig1}
\end{figure}

\begin{figure}[t]%
\includegraphics*[width=\linewidth]{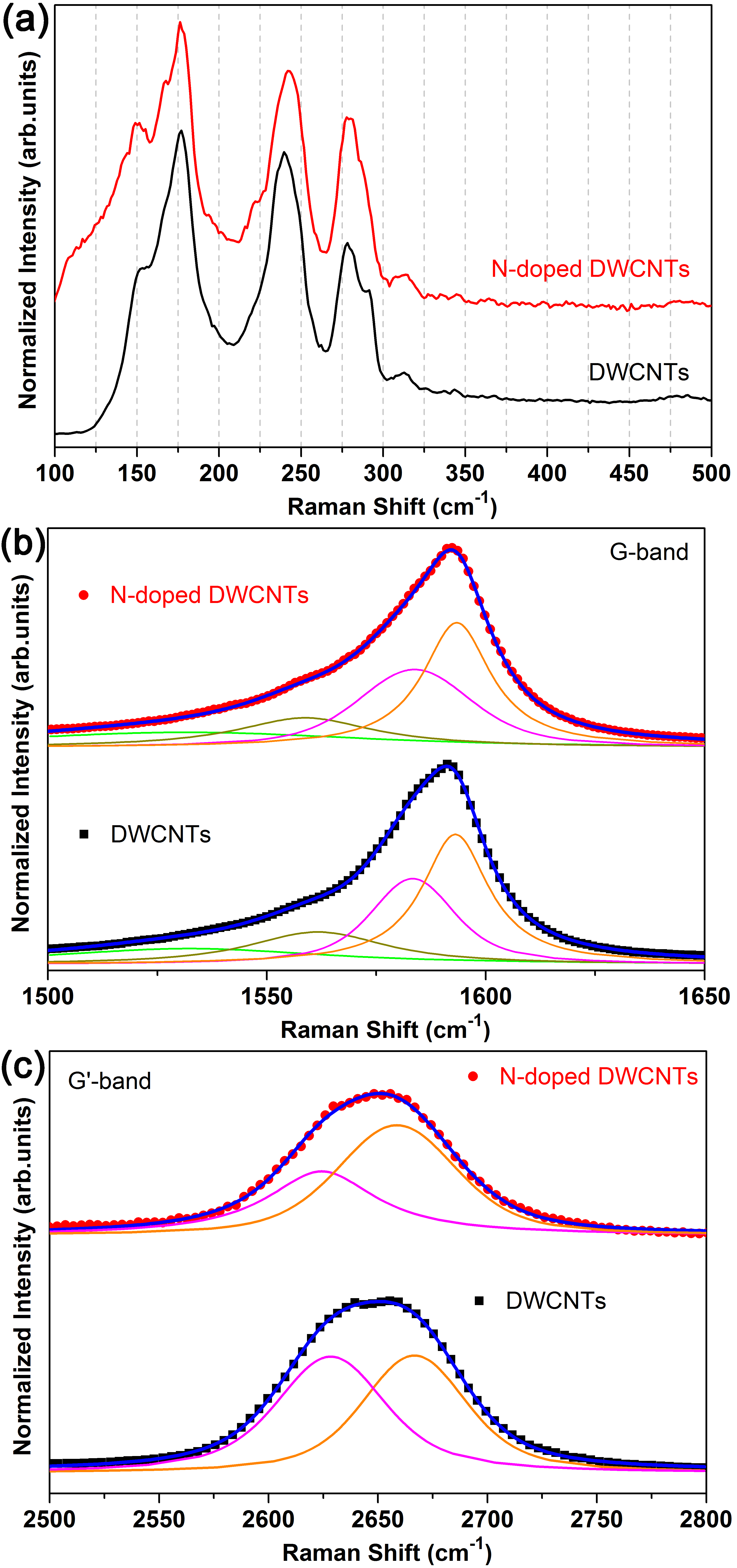}
\caption{%
  ((online color at: www.pss-b.com) Raman spectra of pristine DWCNTs and N-doped DWCNTs excited by 568.188 nm laser. (a) RBM region. (b) G-band. Each G-band has been fitted by four Voigtians. (c) G'-band. Each G'-band has been fitted by two Lorentzians. }
\label{Fig2}
\end{figure}
\section{Results and discussion}
\indent We firstly analyze the Raman spectra of N-doped DWCNTs before and after purification. As seen in Figure 1a, the G/D intensity ratio slightly increases after purification due to the removal of the catalysts, which also indicates that the purification process does not damage the structure of the N-doped CNTs. Some N-doped SWCNTs were also synthesized simultaneously together with N-doped DWCNTs, however, they were almost completely removed by high-temperature treatment in air during the purification process due to their different antioxidation temperatures. Thus, above 95$\%$ of N-doped CNTs in the purified sample are N-doped DWCNTs, which is very similar as the value for the undoped DWCNTs purified by a similar purification process \cite{Shi15arXiv,Rohringer14C,Shi11NR}. Compared to the radial breathing mode (RBM) peaks of the as-grown N-doped DWCNTs, some of the RBM peaks of the purified sample disappear after purification as indicated by the arrows in Fig. 1b, suggesting that these RBM peaks belong to the SWCNTs. From the Kataura plot, we found that the disappeared RBM peaks corresponding to the metallic SWCNTs \cite{Kataura99SM}. In the G-band region, the Breit-Wigner-Fano lineshape for the metallic SWCNTs is normally very wide and intense \cite{Pimenta98PRB,Brown01PRB}. So the diminished G$^-$ after purification was observed (Fig. 1a) because of the population of the metallic SWCNTs decreases.

\begin{figure}[t]%
\includegraphics*[width=\linewidth]{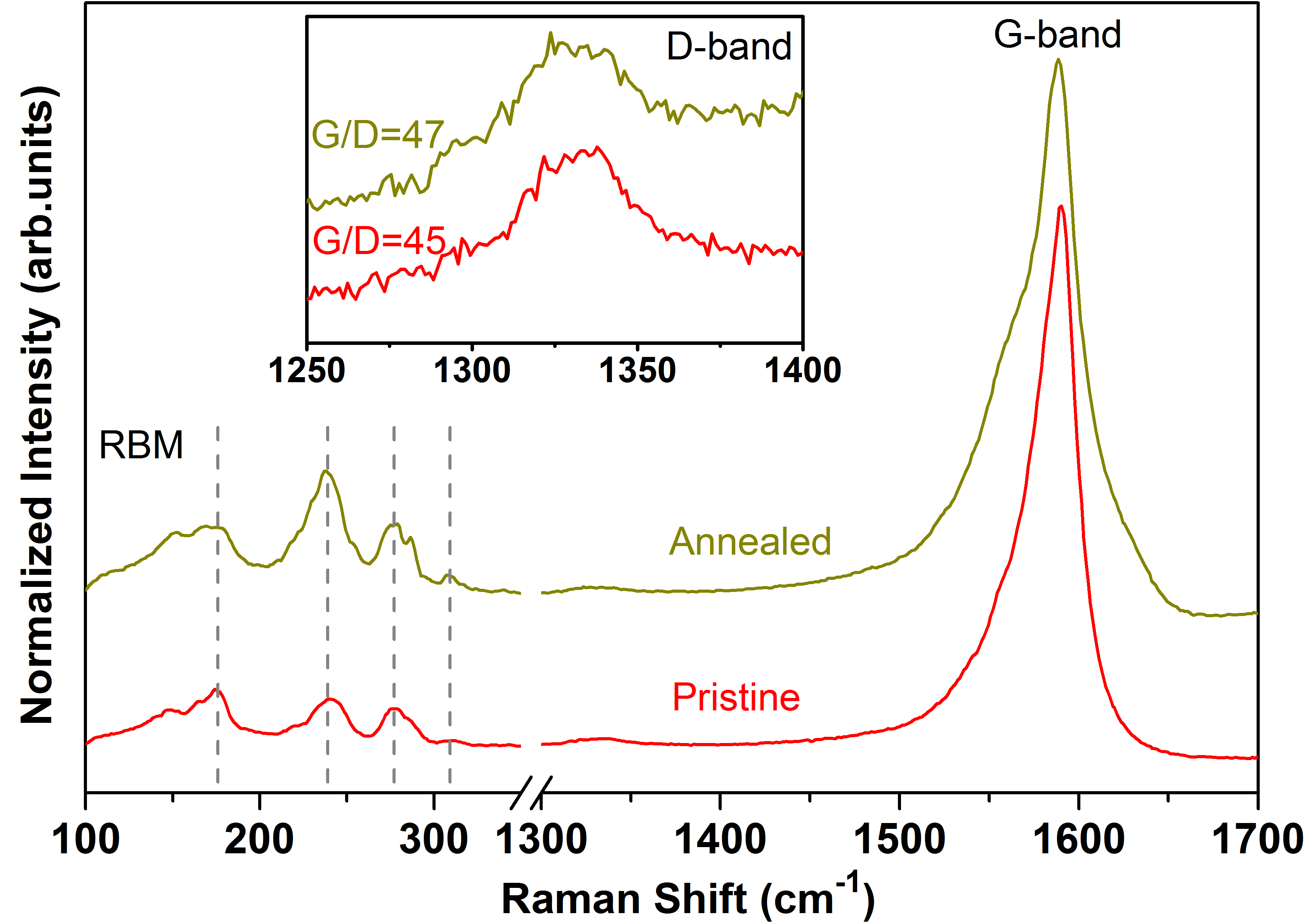}
\caption{%
  ((online color at: www.pss-b.com) Raman spectra of the N-doped DWCNTs before and after annealing in vacuum at the temperature of 1500$\,^{\circ}\mathrm{C}$ excited by 568.188 nm laser. Inset: A close view for the D-band. }
\label{Fig3}
\end{figure}

\indent We didn't observe any change for the RBM peaks due to the doping, as shown in Figure 2a. From the well established inverse proportionality of the radial breathing mode (RBM) to the tube diameter \cite{Kuzmany01EPJB}, we calculated the average diameter of the purified N-doped DWCNTs is around 0.9 and 1.5 nm for the inner and the outer tubes, respectively. This value of the diameter is the same as the undoped DWCNTs \cite{Shi15arXiv}, suggesting that the doping does not change the diameter distribution of DWCNTs synthesized using the same catalysts. This is different from that the thinner N-doped SWCNTs were usually synthesized by introducing the N-dopant \cite{Villalpando-Paez06CPL,Barzegar13TJoPCC}.

\indent The G-band can be fitted by four peaks corresponding to the G$^-$ and G$^+$ of the inner and outer tubes. No Raman shifts of the G$^+$ for both the inner and outer tubes were observed between the doped and undoped DWCNTs (Fig. 2b), because the low-ratio dopants can not change the frequency of the G$^+$ mode \cite{Maciel09PSS(b)}. In addition, the G$^-$ for the doped and undoped tubes are at the same frequency, indicating that the diameter distribution of the doped and the undoped tubes is very similar, because the G$^-$ is highly sensitive to the diameter of the tubes \cite{Jorio02PRB}.

\indent Similar as the G-band behaviour, the G'-band shift was also not observed due to the N-doping. Typically, the G'-band of SWCNTs can be fitted to one Lorentzian \cite{Dresselhaus05PR}. For the DWCNT system, the G'-band can be fitted by two Lorentzians corresponding to the inner and the outer tubes. The overall G'-bands of the undoped and doped samples are at almost the same frequency. However, both fitting peaks for the N-doped DWCNTs were maybe shifted to lower frequencies a little bit (Fig. 2c), which confirms that the doping is N-type \cite{Maciel09PSS(b),Maciel08NM}. Also, the frequency shift is a little bit different for the inner and the outer tubes, suggesting that the doping ratio for the inner and the outer tubes depends on their diameters. This relates with the diameter change of N-doped SWCNTs when doped with different ratio of nitrogen in previous studies \cite{Villalpando-Paez06CPL,Barzegar13TJoPCC}.

\indent In order to test the stability of the N-doped DWCNTs, the sample was annealed at 1500$\,^{\circ}\mathrm{C}$ in vacuum. As shown in Figure 3, compared with the pristine N-doped DWCNTs, the RBM peaks for the annealed ones keep at the same frequencies, which means that the N-doped DWCNTs are very stable, and not destroyed or melted into larger tubes. Theoretically the RBM of high-doping SWCNTs should shift to lower frequency by a few wavenumbers \cite{Gerber09PRB}. However, in our case no frequency shift was observed, indicating that the doping ratio of the N-doped DWCNTs is not high (further discussion in next section). Furthermore, the G/D intensity ratio increases a little bit after annealing, confirming the stability of N-doped DWCNTs again, and also revealing a better crystallinity by annealing. With the same reason, the intensity of RBM peaks of annealed sample is higher than that of the pristine sample as seen in Fig. 3.

\begin{figure}[!t]%
\includegraphics*[width=\linewidth]{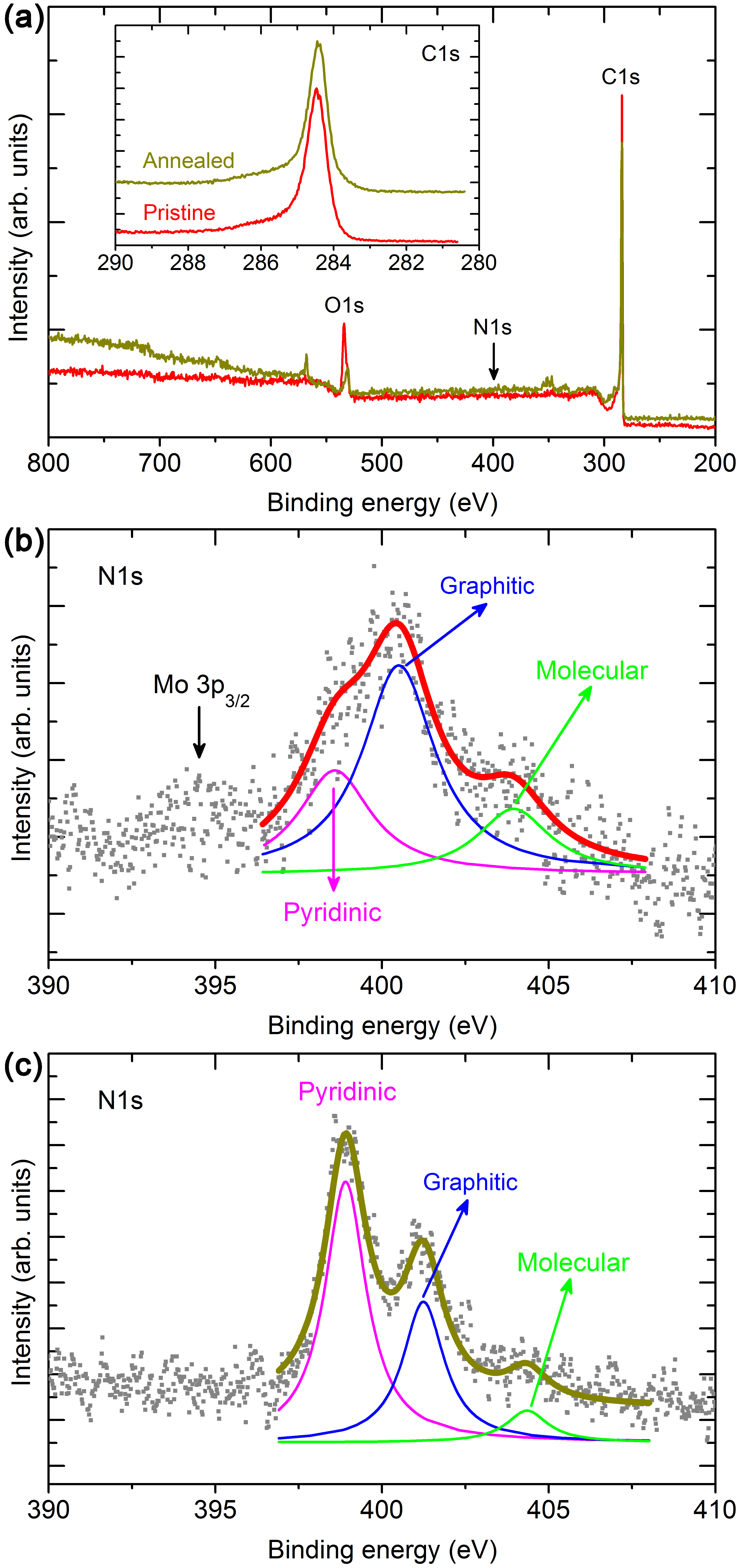}
\caption{%
  ((online color at: www.pss-b.com) XPS of the N-doped DWCNTs before and after annealing. (a) Survey. Inset: A close view for the C1s. (b) N1s of non-annealed N-doped DWCNTs. (c) N1s of annealed N-doped DWCNTs. In (b) and (c), the spectra were fitted by three Voigtians according to different types of N: pyridinic, quaternary, and N-related gas. Note that the pristine and the annealed N-doped DWCNTs is the same sample for XPS measurement.}
\label{Fig4}
\end{figure}

\begin{table}[h]
\caption{Different types of N in pristine and annealed N-doped DWCNTs. The absolute value (relative ratio) for different types of N are shown in at.$\%$ ($\%$).}
\centering
\begin{tabular}{c|c|c|c|c}

  Sample & Pyridinic & quaternary & N-related gas & total\\ \hline
    Pristine & 0.18(27.5) & 0.36(55.3) & 0.11(17.2) & 0.65(100)\\ \hline
    Annealed & 0.66(60.2) & 0.36(32.5) & 0.08(7.3) & 1.10(100)\\
\end{tabular}
\end{table}

\indent Interestingly, the high-temperature annealing not only increases the CNT crystallinity, but also changes the N ratio from 0.65 to 1.10 at.$\%$. As shown in Figure 4b and 4c, there are three types of nitrogen in the N-doped DWCNTs for both of the pristine and the annealed samples: pyridinic, graphitic (or called quaternary, or sp$^2$), and molecular. The results are summarized in Table 1. Firstly, after annealing the molecular nitrogen ratio decreases due to the removal effect by high temperature, but they are not completely removed. It is also confirmed by the largely decreased O ratio after annealing (Fig. 4a) \cite{Ayala10RoMP,Ayala07TJoPCC,Ayala07pss(b)}. Secondly, the graphitic nitrogen ratio does not change at all, indicating that the very good stability of graphitic nitrogen, which can survive under the extremely high temperature in vacuum without problem. Thirdly, the relative content of pyridinic nitrogen is greatly enhanced by annealing from 0.18 to 0.66 at.$\%$. The nitrogen could be from the molecular nitrogen, because it is possible to convert the molecular nitrogen to the pyridinic nitrogen during the annealing by deoxygenation process \cite{Pels95C}. However, apparently the molecular nitrogen is not enough for this large change. Further possibilities will be discussed in next section.

\begin{figure}[!h]%
\includegraphics*[width=0.95\linewidth]{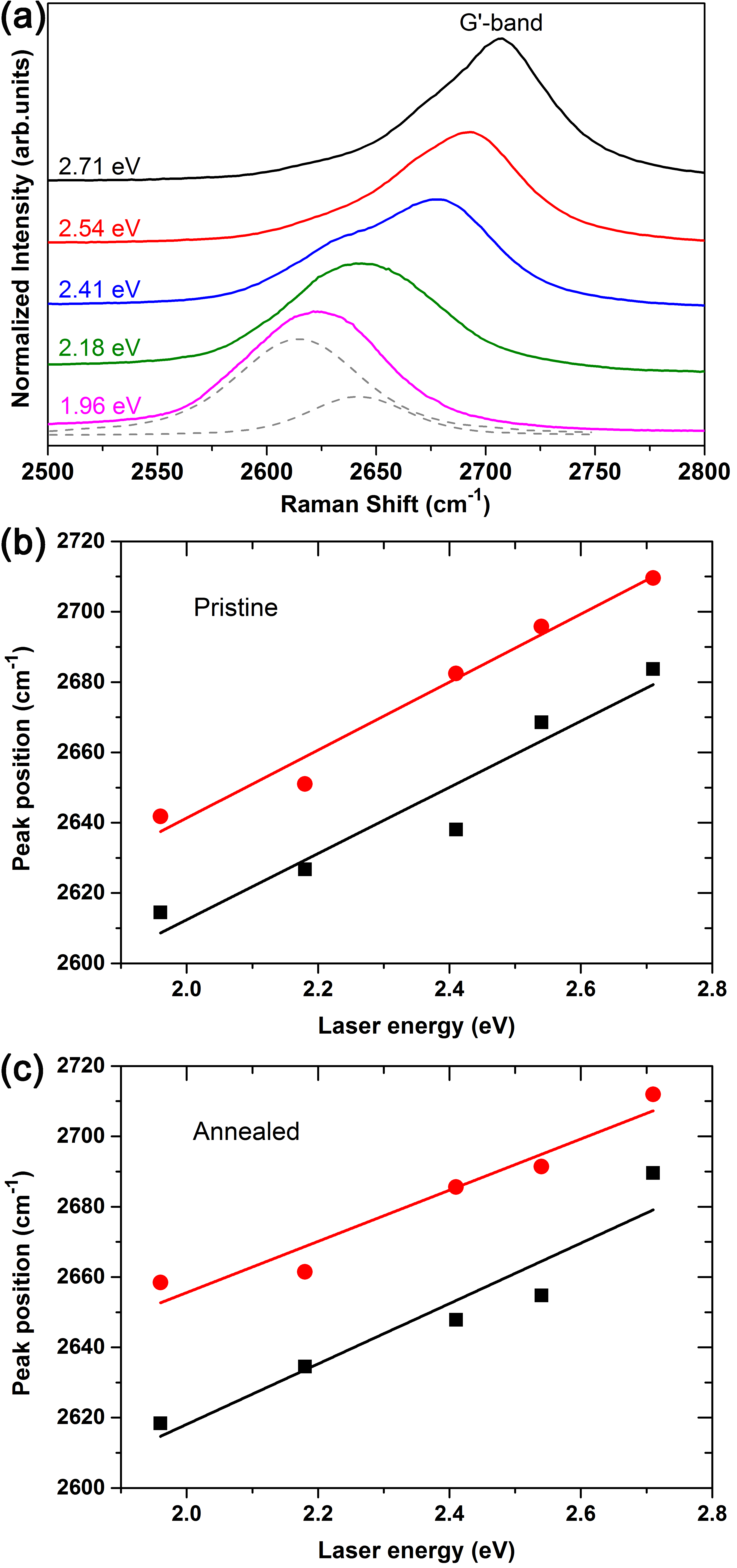}
\caption{%
  ((online color at: www.pss-b.com) (a) The G'-band of N-doped DWCNTs excited by five different lasers. Each G'-band was fitted by two Lorentzians, for example, as shown by the gray dashed lines. The fitted peak positions as a function of excited laser energies for the (b) pristine N-doped DWCNTs and (c) the annealed N-doped DWCNTs. The solid lines show linear line fittings for the experimental data points.}
\label{Fig5}
\end{figure}

\indent Another question is how does the N-dopants affect the G'-band excited by multi-lasers. Previous studies showed that the G'-band of N-doped CNTs almost does not shift after doping. Also, the frequency of G'-band highly depends on the excitation laser energy \cite{Dresselhaus05PR}, and the diameter of the CNTs as well \cite{SouzaFilho03PRB}. We exactly observed these behaviours as shown in Figure 5a. In addition, with all of these knowledge, we would like to analyze the annealing effect on the G'-band of the N-doped DWCNTs. As shown in Fig. 5a, the left and right peaks in the G'-band were assigned to the G'-bands for the inner tubes and the outer tubes, respectively \cite{SouzaFilho03PRB}. A linear fitting was normally used in the studies of the D and G' dispersions \cite{Dresselhaus05PR}. The linear relation can be wrote as $\omega$$_{G'}$ = AE$_{laser}$ + B, where A and B are fitting parameters, and E$_{laser}$ is the excitation laser energy. A higher doping ratio should give out a lower slope \cite{Maciel08NM}. As seen in Fig. 5b and 5c, we obtained very well linear fittings for all the samples, although the slopes are slightly different due to the different doping ratios. Before the annealing, the slopes for the inner and outer tubes ($\omega$$_{G'-inner}$ = (94$\pm$15)E$_{laser}$ + (2424$\pm$37), $\omega$$_{G'-outer}$ = (96$\pm$9)E$_{laser}$ + (2448$\pm$21)) are only slightly lower than the value for the pristine part in the G'-band of N-doped SWCNTs in Ref. \cite{Maciel08NM} ($\omega$$_{G'-pristine}$ = (97$\pm$5)E$_{laser}$ + (2442$\pm$11), $\omega$$_{G'-doped}$ = (71$\pm$4)E$_{laser}$ + (2465$\pm$11)), which means that the doping ratio in our sample is not high. However, after annealing, smaller slopes ($\omega$$_{G'-inner}$ = (86$\pm$15)E$_{laser}$ + (2446$\pm$37), $\omega$$_{G'-outer}$ = (73$\pm$10)E$_{laser}$ + (2510$\pm$25)) were observed as shown in Fig.5c, suggesting that the N ratio increases by the annealing. This is also constant with the XPS results. In addition, we also found that the slope changes for the inner and outer tubes after annealing are different, indicating that the nitrogen ratio for the outer tubes increases more than that for the inner tubes. The N-contained molecules or N-contained amorphous carbon could surround on the DWCNTs, which can be used as N source for further combination into the lattice of the tubes during the annealing, thereby increasing the N ratio. Therefore, larger N ratio increasing for the outer tubes than that for the inner tubes is observed in our experiments. Further investigation should be carefully performed in future. Furthermore, the smaller slope is attributed to the increasing pyridinic nitrogen, because the graphitic nitrogen does not downshift the G'-band \cite{Podila12APL}, and the molecular nitrogen decreases after annealing. That means most of the nitrogen from the N-contained molecules or N-contained amorphous carbon combines into the CNT structure in pyridinic type, but not graphitic type.

\section{Summary}
\indent High-purity N-doped DWCNTs were firstly synthesized from liquid pure benzylamine. The following purification process does not destroy the N-doped DWCNT structure, and the high-temperature annealing in vacuum can increase their crystallinity. The N ratio for the purified N-doped DWCNTs is 0.65 at.$\%$, which is higher than that in previous studies \cite{Ayala07TJoPCC,Ayala07pss(b)}. This doping ratio still can not shift the frequencies of the RBM, the G-band and the G'-band. Three types of N exist in the N-doped DWCNTs: pyridinic, graphitic, and molecular. After annealing, the ratio for the molecular N decreases, no change for the graphitic N, and increasing a lot for the pyridinic N. This great increasing is mostly contributed from the higher-doped outer tubes due to the combination of the surrounded N-contained materials into the structure of the DWCNTs. \\

\begin{acknowledgement}
This work was supported by the Austrian Science Funds (FWF, NanoBlends I 943--N19). L. S. thanks the scholarship supported by the China Scholarship Council (CSC). P. A. was supported by a Marie Curie Intra European Fellowship within the 7th European Community Framework Program.
\end{acknowledgement}

%
%


\bibliographystyle{pss}
\bibliography{IWEPNM2015}


\end{document}